\documentstyle[11pt,adassconf]{article}  

\begin{document}   

%

\paperID{D-12}


\title{Hands-On TAROT: Intercontinental use of the TAROT for Education and Public Outreach}

\author{Michel Bo\"er, Carole Thi\'ebaut, Alain Klotz}
\affil{Centre d'Etude Spatiale des Rayonnements (CESR/CNRS), BP
4346, F 31028 Toulouse Cedex 4 France}
\author{G. Buchholtz}
\affil{Institut National des Sciences de l'Univers, Division
Technique (CNRS), Meudon, France }
\author{A.-L. Melchior}
\affil{DEMIRM/CNRS, Observatoire de Paris, France}
\author{C. Pennypaker, M. Isaac}
\affil{University of California at Berkeley, USA}
\author{Toshikazu Ebisuzaki}
\affil{RIKEN, Tokyo, Japan}

\contact{Michel Boer}
\email{Michel.Boer@cesr.fr }

\paindex{Boer, M.} \aindex{Thi\'ebaut, C.} \aindex{Buchholtz, G.}
\aindex{Melchior, A.-L.} \aindex{Pennypaker, C.} \aindex{Isaac,
M.} \aindex{Ebisuzaki, T.}



\begin{abstract}          

The TAROT telescope has for primary goal the search for the prompt
optical counterpart of Cosmic Gamma-Ray Bursts. It is a completely
autonomous 25cm telescope installed near Nice (France), able to
point any location of the sky within 1-2 seconds. The control,
scheduling, and data processing activities are completely
automated, so the instrument is completely autonomous. In
addition to its un-manned modes, we added recently the
possibility to remotely control the telescope, as a request of
the "Hands-On Universe" (HOU) program for exchange of time within
automatic telescopes for the education and public outreach. To
this purpose we developed a simple control interface. A webcam
was installed to visualize the telescope. Access to the data is
possible through a web interface. The images can be processed by
the HOU software, a program specially suited for use within the
classroom. We experienced these feature during the open days of
the University of California Berkeley and the Astronomy Festival
of Fleurance (France). We plan a regular use for an astronomy
course of the Museum of Tokyo, as well as for French schools. Not
only does Hands-On TAROT gives the general public an access to
professional astronomy, but it is also a more general tool to
demonstrate the use of a complex automated system, the techniques
of data processing and automation. Last but not least, through
the use of telescopes located in many countries over the globe, a
form of powerful and genuine cooperation between teachers and
children from various countries is promoted, with a clear
educational goal.

\end{abstract}


\section{Introduction}

They have been several attempts of using astronomical data in the
classroom, in general within the framework of physics,
mathematics, and/or astronomy courses. Using directly a telescope
in the college backyard has many advantages, mainly that children
themselves practice  astronomy with a telescope. However, several
problems may arise:

\begin{itemize}

\item Except for the Sun, astronomical observations take place
at night, making them somewhat difficult to accommodate on a
regular basis, both for pupils and teachers.

\item Many schools are located in town, and do not have any
dark area where to locate a telescope at night.

\item Teachers are typically not experienced astronomers.

\item Having a telescope in the school requires some care in
handling and maintaining it.

\item Not all colleges can afford a telescope with (or even
without) a CCD camera.

\end{itemize}

To that purpose, the Hand-On Universe program (Pennypaker et al.
1998; Bo\"er et al. 2001) has been initiated to use astronomical
data within the classroom. Telescope time is exchanged within the
HOU network, in order to enable the use various telescopes over
the world. Most of them may be remotely controlled, allowing to
use them at night.

\section{TAROT, an autonomous observatory}

The prime objective of the {\it T\'elescope \`a Action Rapide pour
les Objets Transitoires} (TAROT; Bo\"er et al., 1999; Bo\"er et
al. 2000; \htmladdnormallink{http://tarot.cesr.fr} )), is the real
time observation of cosmic Gamma-Ray Bursts (hereafter GRBs).
TAROT is a 25cm telescope, with a full autonomous control system,
and able to point any location over the sky within 1-2 seconds.
Figure 1 displays the functional diagram of TAROT. In normal
operations, the {\it MAJORDOME} (Bringer et al., 2000) computes
the schedule and sends observation requests to the {\it Telescope
Control System}, which takes care of the various housekeeping,
points the telescope, and activates the CCD Camera. As soon as
the data is taken, it is pre-processed, with dark, bias and
flat-field substraction, cosmic ray removal, astrometric
reduction, and a source list is built. The requests for
observations are now sent via the web. Should a GRB alert occurs
(from the HETE-2 satellite), the present observation is
interrupted, and the telescope slews immediately to the position
of the GRB source.


%



The interfaces with the users, beside the "alert" connection with
the GCN, are  as follows:

\begin{itemize}

\item The main interface has now been rewritten as a web form. The
user is requested to write the coordinate of the source,  the
number (up to 6), duration and filter(s) (6 positions) of the
frames he/she wants. A unique identifier is attributed to the
request as soon as the user validates it. The new request is taken
into account by the {\it MAJORDOME} software at the next start of
the scheduling process, at least once a day or whenever some event
interrupts the operation at night, e.g. rain. Since the form is
available through the web, the user can use any computer system.

\item As a request of the HOU program, we included a direct remote
interface, written in the Java language, again to avoid any
preference to a particular operating system. The user can operate
directly the telescope, provided the {\it MAJORDOME} accepts
input from this interface. Any kind of image can be acquired from
this interface. However, one of the telescope operators has to
log in the system to allow operations through this interface.

\end{itemize}





\section{Discussion}

We tested the various TAROT user interfaces at several public
demonstrations. They proved to be very reliable. During the day,
the presence of a webcam enables the user to see the immediate
reaction of an instrument located at several hundreds or thousands
kilometers from him. At night, images are available through the
web within one or two minutes, on a page which includes the image
in jpeg, and the fits header. Optionally, the sources from the
USNO A2.0 catalog can be superimposed on the image (Thi\'ebaut
and Bo\"er, 2001), an asteroid chart can be requested, and the DSS
can be extracted using a preformatted  SKYVIEW query.

Since the prime goal of TAROT is doing science, we still prefer
that users from schools either use frames from the scientific
program (including frames acquired during the last night), or
send requests to the batch interface, reserving the direct remote
interface for demonstration purposes during the day or at night.
We plan also to enhance this interface par allowing the {\it
MAJORDOME} to schedule in advance the blocks of nights allowed
for a use in direct access mode.

We found also that what seems evident to the astronomer, has to
be explained to general audiences, e.g. phenomena like saturation
of frames, angles expressed in hours, and that a telescope
located in the northern hemisphere has some difficulties to look
at e.g. the Magellanic Clouds (this has also to be explained to
several astronomers), or that the accessible sources in the sky
varies from winter to summer. To cope with these last points, we
plan to have a more interactive and pedagogical interface. In any
case, this exercise of porting a system devoted to a somewhat
specialized audience to the general public proved to be a very
interesting and rewarding adventure for the TAROT team.

\acknowledgments

The TAROT program is funded by the Centre National de la
Recherche Scientifique, Institut National des Sciences de
l'Univers (CNRS/INSU).


%


\end{document}